\documentclass[a4paper,11pt]{article}
\usepackage{pos}

\newcommand\beq{ \begin{eqnarray} }
\newcommand\eeq{ \end{eqnarray} }

\usepackage{color}

\title{QCD viscosity by combining the gradient flow and sparse modeling methods}

\author*[a,b,c,d]{Etsuko Itou}
\affiliation[a]{Strangeness Nuclear Physics Laboratory,
RIKEN Nishina Center, Wako 351-0198, Japan}
\affiliation[b]{Interdisciplinary Theoretical and Mathematical Sciences Program (iTHEMS), RIKEN, Wako 351-0198, Japan}
\affiliation[c]{Department of Physics, and Research and 
Education Center for Natural Sciences, Keio University, 4-1-1 Hiyoshi, Yokohama, Kanagawa 223-8521, Japan}
\affiliation[d]{Research Center for Nuclear Physics (RCNP), Osaka University, Osaka 567-0047, Japan}
\emailAdd{itou@yukawa.kyoto-u.ac.jp}

\author[e,f]{Yuki Nagai}
\affiliation[e]{
CCSE, Japan  Atomic Energy Agency, 178-4-4, Wakashiba, Kashiwa, Chiba, 277-0871, Japan}
\affiliation[f]{
Mathematical Science Team, RIKEN Center for Advanced Intelligence Project (AIP), 1-4-1 Nihonbashi, Chuo-ku, Tokyo 103-0027, Japan
}
\emailAdd{nagai.yuki@jaea.go.jp}

\abstract{We give a new description to obtain the shear viscosity in QCD at finite temperature.
Firstly, we obtain the correlation function of the renormalized energy-momentum tensor using the gradient flow method.
Secondly, we estimate the spectral function from the smeared correlation functions using the sparse modeling method.
The combination of these two methods looks promising to determine the shear viscosity precisely.}

\FullConference{%
 The 38th International Symposium on Lattice Field Theory, LATTICE2021
  26th-30th July, 2021
  Zoom/Gather@Massachusetts Institute of Technology
}

\begin{document}
\maketitle

\section{Introduction and Summary}
Shear viscosity in QCD is one of the most important physical quantities to understand the properties of QCD.
In particular, RHIC experiments and hydrodynamic analysis suggest that the shear viscosity to entropy ratio, $\eta/s$, is very small around the critical temperature ($T_c$), which indicates a perfect liquid of quarks and gluons. On the other hand, the first-principles calculation of $\eta/s$ has some inherent difficulties, and its precise determination and its temperature dependence have not yet been achieved.
\begin{figure}[h]
\begin{center}
\includegraphics[scale=0.4]{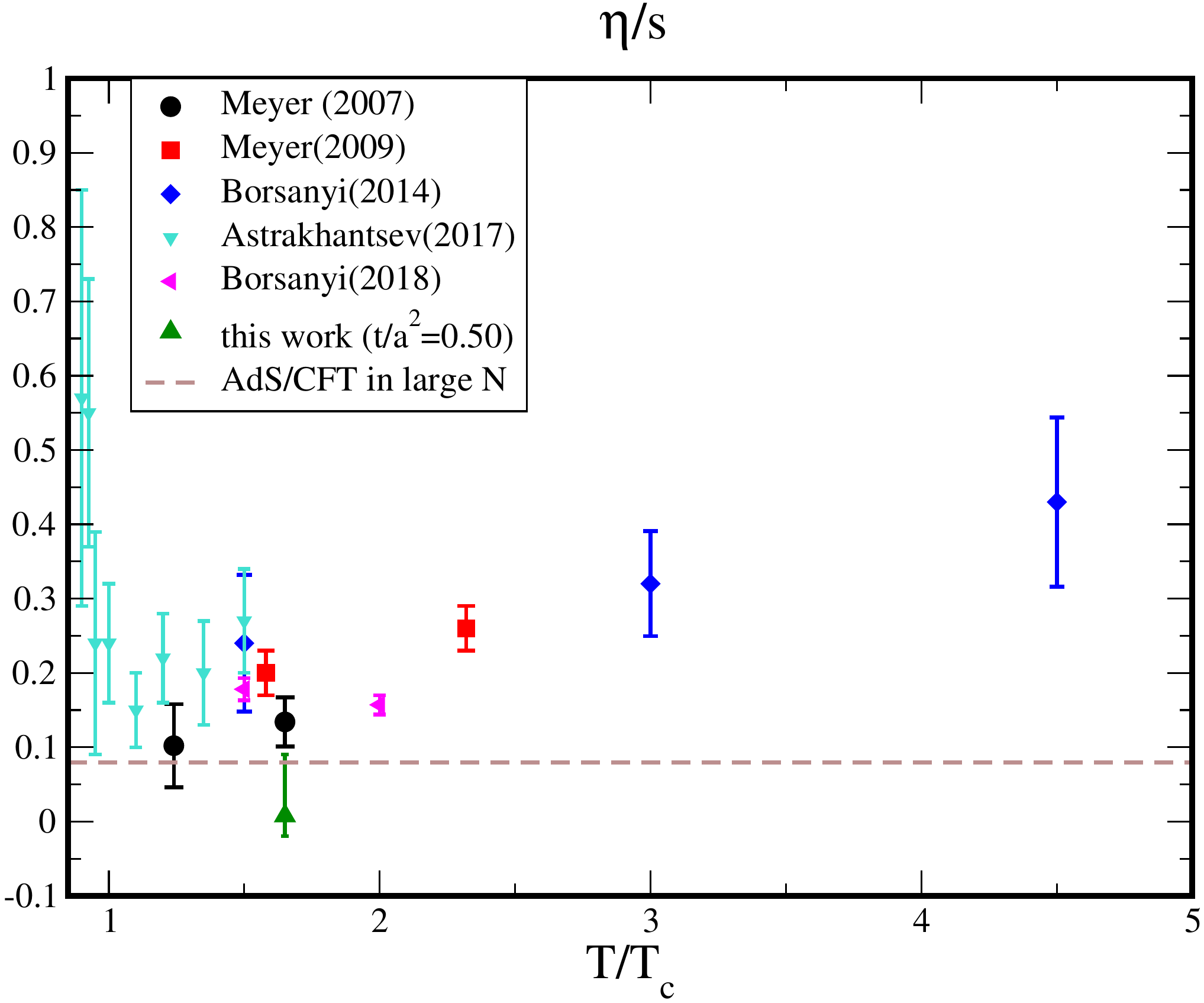}
\caption{Current status on the $T$-dependence of $\eta/s$ in quenched QCD.
Our result obtained in Ref.~\cite{Itou:2020azb} is plotted as a green symbol at $T/T_c=1.65$.
The other symbols have given by Refs.~\cite{Meyer:2007ic, Meyer:2009jp, Mages:2015rea, Astrakhantsev:2017nrs,Pasztor:2018yae}.
}
\label{fig:comp-eta-s}
\end{center}
\end{figure}

The shear viscosity is given by the first-differential coefficient of the spectral function at zero frequency ($\eta (T) = \pi d \rho (\omega)/ d \omega $ at $\omega=0$).
The spectral function is defined by the Euclidean correlation function of the renormalized spacial energy-momentum tensor (EMT) in the static state as follows:
\beq
C(\tau) &=& \frac{1}{T^5} \int d \vec{x} \langle T^R_{12} (0, \vec{0}) T^R_{12}(\tau, \vec{x}) \rangle= \int^{+\infty}_{-\infty} d \omega K(\tau, \omega) \rho(\omega).\label{eq:fisrt-eq}
\eeq 
Here, $K(\tau, \omega) = \frac{\cosh \left( \omega (\frac{1}{2T}-\tau) \right)}{\sinh (\frac{\omega}{2T})}$, 
denotes the kernel of an integral transform.
In the lattice simulations, $C(\tau)$ is measured by generated configurations using the first equality sign, and then $\rho(\omega)$ is estimated through the second one.

During these calculations, there are at least three essential difficulties to obtain $\rho(\omega)$:
\begin{enumerate}
    \item  How to define the renormalized EMT on the lattice
    \item How to improve a bad signal-to-noise ratio of the correlation function of EMT
    \item How to estimate $\rho(\omega)$ from the limited number of the data $C(\tau)$ 
\end{enumerate}
As for the first and second difficulties, the gradient flow method~\cite{Suzuki:2013gza,Asakawa:2013laa} looks promising.
Thanks to the UV finiteness of the flowed operators in the gradient flow~\cite{Luscher:2010iy,Luscher:2011bx},
we can define a renormalized EMT without hard numerical calculation of $Z$-factor for quenched QCD.
Furthermore, the gradient flow is equivalent to a continuum stout smearing, so that it reduces the statistical uncertainty.
Actually, in our work~\cite{Itou:2020azb} the number of gauge configurations for the measurements is only $2,000$. 
It is quite a small number of configurations in comparison with former works; $800,000$ in Ref.~\cite{Nakamura:2004sy} and $6$ million in Ref.~\cite{Pasztor:2018yae}.

On the other hand, we find that the utilizing the gradient flow makes the third problem harder.
The third problem is well known as an ill-posed inverse problem. 
Especially in the finite temperature system, the number of sites in the temporal direction is very limited, then it makes harder to solve the inverse problem.
If we use the gradient flow method to obtain the correlation function, we have to take the limited data in $\tau > \sqrt{8t}$, where $\sqrt{8t}$ denotes the smeared regime at flow-time $t$. 
If the distance of two operators ($\tau$) gets shorter than the smeared length ($\sqrt{8t}$) by the gradient flow, then the smeared regime of one operator reaches at the position of the other operator in the correlation function.
We cannot measure a correct correlation between them in such situations.
Therefore, the available number of the data is further limited if we utilize the gradient flow method to obtain $C(\tau)$.

In this work, we propose the sparse modeling method~\cite{Shinaoka2017a, Shinaoka2017b} to estimate the spectral function at finite flow-time~\footnote{The sparse modeling method can be applied to both non-smeared and smeared correlation functions.}.
Utilizing the intermediate-representation basis (IR basis) and a few reasonable constraints, the sparse modeling makes it possible to estimate the real-frequency representation of $\rho(\omega)$ from a sparse imaginary-time correlation function.

The status on $\eta/s$ in quenched QCD is shown in Fig.~\ref{fig:comp-eta-s}.
Here, our result obtained in Ref.~\cite{Itou:2020azb} is plotted as a green symbol at $T/T_c=1.65$, where we have not yet taken the continuum limit and $t\rightarrow 0$ limit.
Actually, some data in Fig.~\ref{fig:comp-eta-s} are the results after taking continuum limit, while the other ones are not.
From the figure, we can see a general trend that $\eta/s$ decreases with temperature and 
it approaches the Kovtun-Son-Starinet (KSS) bound ($\eta/s=1/(4\pi)$) around $T=T_c$ which is the lower bound for $\eta/s$ predicted by the large $N_c$ analysis based on AdS/CFT correspondence~\cite{Kovtun:2003wp,Son:2007vk}.

\section{Measurement of the correlation function of EMT in the gradient flow method}\label{sec:EMT}
The renormalized EMT for the quenched QCD in the gradient flow method~\cite{Suzuki:2013gza} is given by 
\begin{align}
   T_{\mu\nu}^R(x)
   =\lim_{t\to0}\left\{\frac{1}{\alpha_U(t)}U_{\mu\nu}(t,x)
   +\frac{\delta_{\mu\nu}}{4\alpha_E(t)}
   \left[E(t,x)-\left\langle E(t,x)\right\rangle_0 \right]\right\},
\label{eq:(4)}
\end{align}
at a flow-time $t$.
Here, we utilize the small flow-time expansion and drop ${\cal O} (t)$ corrections.
The coefficients $\alpha_U,\alpha_E$ are calculated perturbatively in Ref.~\cite{Suzuki:2013gza}.
An advantage of the usage of the gradient flow method is the absence of $Z$-factor to define the renormalized EMT.
Once we use the renormalized coupling constant in the coefficient $\alpha_U, \alpha_E$, all composite operators constructed by the flowed gauge field are the UV finite at positive flow-time ($t>0$)~\cite{Luscher:2011bx}.
The last term in right-hand-side, $\langle\cdot\rangle_0$, describes the vacuum expectation value (v.e.v.), which relates to the zero point energy. 
$U_{\mu\nu}(t,x)\equiv G_{\mu\rho}(t,x)G_{\nu\rho}(t,x)
-\delta_{\mu\nu} E(t,x)$
and~$E(t,x)\equiv\frac{1}{4}G_{\mu\nu}(t,x)G_{\mu\nu}(t,x)$~\footnote{Here, we drop the color indices.} denote gauge-invariant local products of dimension~$4$.
 $G_{\mu \nu}$ represents the field strength constructed by the flowed gauge field ($B_{\mu}(t,x)$), and on the lattice it can be calculated by the clover-leaf operator.

Now, we calculate the two-point function of EMT.
The shear viscosity is given by the Euclidean correlation function of $T^R_{12}$ component defined as
\beq
C(\tau) = \frac{1}{T^5} \int d \vec{x} \langle T^R_{12} (0) T^R_{12}(x) \rangle,\label{eq:def-Ctau-latt}
\eeq 
where $x=(\tau, \vec{x})$.
On the lattice at a finite flow-time $t$, we measure the two-point function of $U_{12} (t,x)$,
\beq
C (t,\tau/a)= N_\tau^5 \sum_{\vec{x}} \frac{1}{\alpha_U(t)^2} \langle U_{12} (t, 0) U_{12}(t,x) \rangle,\label{eq:def-smear-C}
\eeq
where the second argument of $U_{12} (t,x)$ denotes a four-dimensional vector $x=(\tau/a, \vec{x}/a)$.


\section{Sparse modeling method}\label{sec:sparse-modeling}
Here, we give a brief review of the sparse modeling method in the IR basis following Refs.~\cite{Shinaoka2017a, Shinaoka2017b} (see also a review paper~\cite{review-sparse}). 

The input is the imaginary-time Green's function $C(\tau)$.
Its Fourier transform $C(i \omega_n)$, where $\omega_n$ denotes a Matsubara frequency, is related with 
the spectral function $\rho(\omega)$ as $\rho(\omega)= \frac{1}{\pi} \mbox{Im} C(\omega + i 0)$ by replacing $i \omega_n$ with $\omega + i0$.
Then, the input $C(\tau)$ is written by the integral form of the spectral function,
\beq
C(\tau) = \int^{+\infty}_{-\infty} d \omega K(\tau, \omega) \rho(\omega), \label{eq:def-C}
\eeq
where $0 \le \tau \le 1/T$.

For simplicity,  we denote Eq.(\ref{eq:def-C}) using the dimensionless vectors as
\beq
\vec{C} \equiv K \vec{\rho}. \label{eq:ckrho}
\eeq
The component of the vector $\vec{C}$ is $C_i \equiv C(\tau_i)$, where $\tau_i$ labels the temporal site on the lattice with $0 \le \tau_i \le N_\tau-1$.
The kernel $K$ becomes $N_\tau \times N_\omega$ matrix, while $\vec{\rho}$ denotes a vector whose component is $\rho_j \equiv \rho(\omega_j)$ with $j = 1, \cdots ,N_\omega$.
Generally, if $N_\omega > N_\tau$, then we cannot find an unique solution for $\vec{\rho}$.
That is an essential difficulty of the inverse problem.

To solve this problem, we change the basis if we find the solution.
By introducing the singular value decomposition (SVD) of the matrix $K$ defined as 
\beq
K = U S V^\dag,
\eeq
we rewrite Eq.(\ref{eq:ckrho}) as 
\beq
\vec{C} \equiv  U S V^\dag \vec{\rho}.
\eeq
Here $S$ is an $N_\tau \times N_\omega$ diagonal matrix, and $U$ and $V$ are unitary matrices of size $N_\tau \times N_\tau$ and $N_\omega \times N_\omega$, respectively.

Now, we take new vectors in the IR basis (SVD basis),
\beq
\vec{\rho}' \equiv V^t \vec{\rho}, ~~~~~~~~ \vec{C}' \equiv U^t \vec{C}.\label{eq:rho-C-IR-basis}
\eeq
The new vectors $S\vec{\rho}'$ and $\vec{C}'$ are represented as the same $N_\tau$-component vector.

In this work, we find $\vec{\rho}$, which minimize the following cost function,
\beq
F(\vec{\rho}') \equiv \frac{1}{2} \| \vec{C}' - S \vec{\rho}'  \|_2^2 + \lambda \| \vec{\rho}' \|_1,\label{eq:cost-fn-L1}
\eeq 
under the non-negative condition for the spectral function.
Here, we introduce the  $L_1$ regularization term, $\lambda \| \vec{\rho}' \|_1$, where $\lambda$ denotes a Lagrange multiplier.
This regularization term reduces the number of components of $\vec{\rho}'$.
If we choose a optimized value of $\lambda$, then it makes the optimization problem above stable against noise or statistical fluctuation.

The numerical algorithm to solve the optimization problems with these constraints has been developed~\cite{ADMM1, ADMM2}, and it is called the alternating direction method of multipliers (ADMM) algorithm.
See Appendix~A in Ref.~\cite{Itou:2020azb} for the detail of the algorithm.

\section{Strategy to obtain the shear viscosity  using the sparse modeling method}\label{sec:sparse-Ctau}
Here, we give a standard formula for the sparse modeling method.
In the actual analysis of our work, although we modify the standard formula for applying the smeared correlation function at a finite flow-time, the processes to find the spectral function from the correlation function is the same with the standard one.
The numerical code and the data are available on the arXiv page~\cite{Itou:2020azb}.

The integral equation which we would to solve is given by
\beq
C(\hat{\tau})&=& N_\tau^5 \int^{\infty}_0 d \hat{\omega}  \hat{\rho}( \hat{\omega}) \frac{\cosh \hat{\omega} \left( \frac{N_\tau}{2} -\hat{\tau} \right)}{\sinh (\frac{\hat{\omega} N_\tau }{2})},\label{eq:EMT-corr-org}
\eeq
where $\hat{\tau}=\tau/a$, $\hat{\omega}=a \omega$ and $\hat{\rho} = a^4 \rho $, respectively.
We normalize the correction function using $C(\hat{\tau}=0)$ in the calculation basis, 
\beq
C_{calc} (\tau_i') &=& \frac{C(\tau'_i)}{N_\tau^4 \Lambda} \left[  \frac{C(\tau'_0)}{N_\tau^4 \Lambda  \sqrt{\Delta \omega'}}   \right]^{-1}= \frac{C(\tau_i')}{C(\tau'_0)} \sqrt{\Delta \omega '}.\label{eq:norm-C}
\eeq
Here, $\omega'= \hat{\omega}/\hat{\omega}_{cut}, \tau' = \hat{\tau}/N_\tau$.
We also introduce the spectral function in the calculation basis ($\tilde{\rho}_{calc}$),
\beq
\tilde{\rho} (\omega_j ')  = \frac{\hat{\rho} (\hat{\omega})}{2\tanh (\hat{\omega}N_\tau/2)}
= \frac{C(\tau'_0)}{N_\tau^4 \Lambda \Delta \omega '} \tilde{\rho}_{calc} (\omega_j ').\label{eq:tilde-rho-calc}
\eeq
Then the discretized kernel is represented as 
\beq
K(\tau'_i,\omega'_j)  \equiv \frac{\cosh [\frac{\omega_j ' \Lambda }{2} (2 \tau_i' -1)]}{\cosh (\omega_j ' \Lambda /2)} \sqrt{\Delta \omega'}.\label{eq:kernel-SVD}
\eeq
We carry out SVD decompositionof this discretized kernel. Here, we utilize DGESDD routines of LAPACK.

Next, we estimate an optimal value of $\lambda$ using ADMM algorithm and find the optimized value of $\lambda$, which fixes the strength of the $L_1$ regularization term.
Finally, we carry out the sparse modeling analysis using the ADMM algorithm to find the most likely spectral function $\tilde{\rho}_{calc}$ using the optimized value of $\lambda$.
The shear viscosity divided by $T^3$ to be a dimensionless quantity is given by 
\beq
\frac{\eta}{T^3} &\equiv & \pi \frac{1}{T^3} \frac{d \rho (\omega)}{d \omega} |_{\omega =0}, \nonumber\\
&=& \pi {N_\tau^4} \tilde{\rho} (0). \label{eq:eta-rho-calc}
\eeq

As a check, we reconstruct the correlation function using the obtained $\tilde{\rho}_{calc}$ as 
\beq
C_{output}(\tau'_i) = \frac{C(\tau'_0)}{\sqrt{\Delta \omega}} \sum_j K(\tau'_i, \omega'_i) \tilde{\rho}_{calc} (\omega'_i) .
\eeq
It is a feasibility test of the analysis whether  $C_{output} (\tau_i)$ reproduces the input correlation function $C(\tau_i)$.

\section{Simulation detail}
\subsection{Simulation setup}\label{sec:setup}
We utilize the Wilson plaquette gauge action under the periodic boundary condition at $\beta=6.93$ on $N_s^3 \times N_\tau = 64^3 \times 16$ lattices.
The lattice parameter realizes the system at $T=1.65T_c$. 
We use the relation between $a/r_0$, where $r_0$ denotes the Sommer scale, and $\beta$ in Ref.~\cite{Guagnelli:1998ud}.
Then,  $T/T_c$ is fixed by the resultant values of $Tr_0 = (N_\tau (a/r_0))^{-1}$ using the result at $\beta= 6.20$ in Ref.~\cite{Boyd:1996bx}.
 The gradient flow method in Ref.~\cite{Asakawa:2013laa} gives the thermal entropy, $s/T^3=4.98(24)$, after $a \rightarrow 0$ and then $t\rightarrow 0$ extrapolations.

Gauge configurations are generated by the pseudo-heatbath algorithm with the over-relaxation.
We call one pseudo-heatbath update sweep plus several over-relaxation sweeps
as a ``Sweep". 
To eliminate the autocorrelation, we take $200$ Sweeps between measurements. 
The number of gauge configurations for the measurements is $2,000$. 
As for the gradient flow, we utilize the third-order Runge-Kutta method with $\varepsilon= 0.01$, and confirm that the accumulation errors are sufficiently smaller than the statistical errors.
The gauge action of the flow is the Wilson plaquette gauge action.

To estimate  statistical uncertainty of the spectral function, we perform the bootstrap analyses.
We resample $N_{boot}$ sets of $2,000$ data of $C(t,\tau/a)$ for each configuration, where the overlapping selection of the configurations for one bootstrap sample is allowed.
For each set of the bootstrap sample, we take the average of $2,000$ data of $C(t, \tau/a)$ and carry out the sparse modeling analysis using its mean value.
The statistical errors of the spectral function are calculated by the variance over $N_{boot}$ samples.
Here, we take $N_{boot}=1,000$.
We also discussed several systematic errors in the sparse modeling analysis in Ref.~\cite{Itou:2020azb}.

\subsection{Simulation results }\label{sec:stat-error}

\begin{figure}[h]
\begin{center}
\includegraphics[scale=0.45]{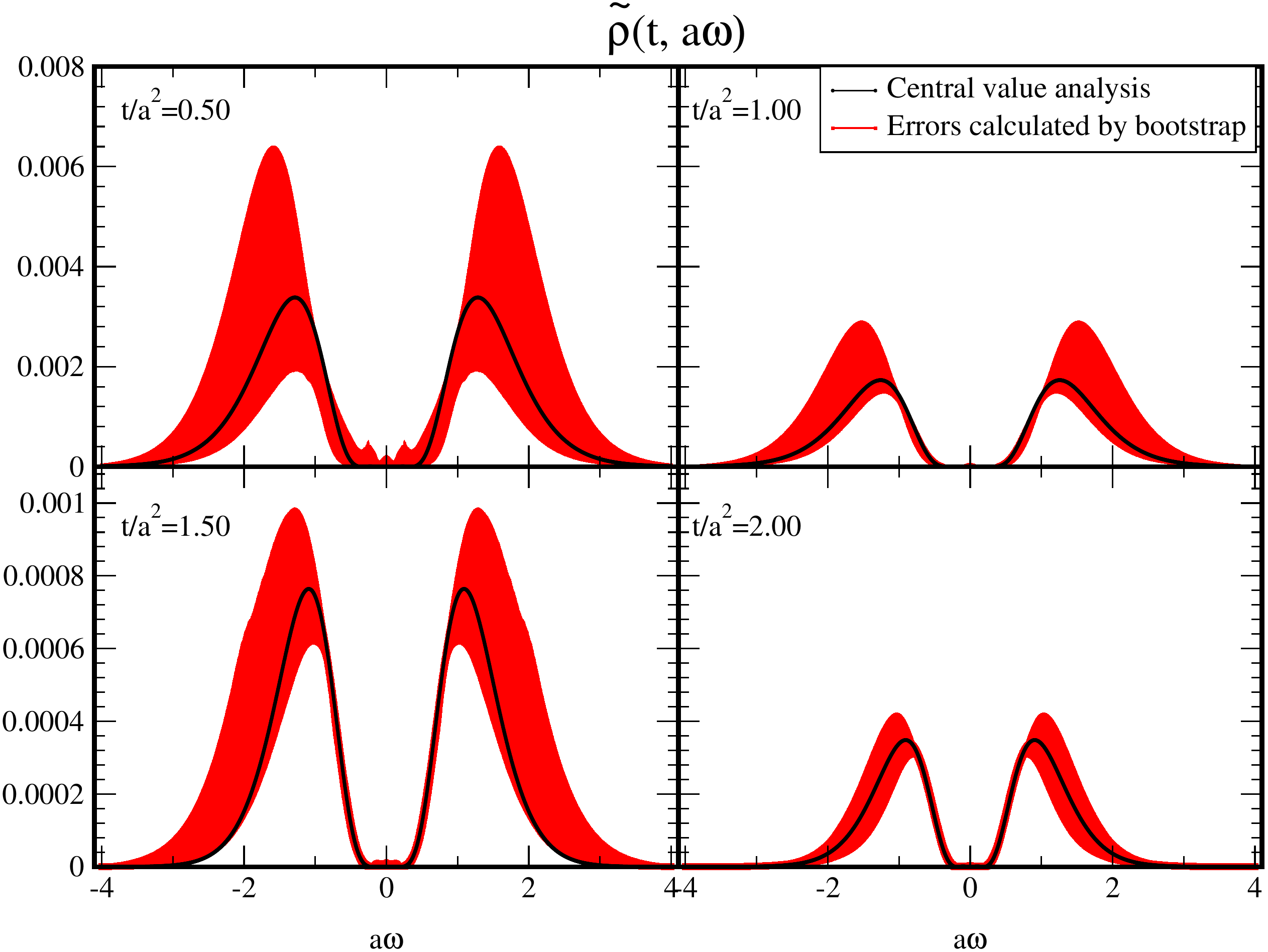}
\caption{The statistical error of the spectral function using the bootstrap analysis. }
\label{fig:rho-errors}
\end{center}
\end{figure}
Figure~\ref{fig:rho-errors} depicts the obtained spectral function with the bootstrap errors.
We also plot the result using the central-value of $C(t,\tau/a)$ as black curve.

We see the integration of obtained $\tilde{\rho}(t,a\omega)$ in terms of $a \omega$, 
\beq
{\mathcal N}=\int_{-a\omega_{cut}}^{a\omega_{cut}} \tilde{\rho}(t,a\omega) d (a\omega),\label{eq:N}
\eeq
monotonically decreases as a function of flow-time.
In more detail, the spectral functions in large $|\omega|$ regime are highly suppressed in the longer flow-time.
The gradient flow gradually reduces the degree of freedom with high-frequency and can be interpreted as a renormalization group flow.
The results of the sparse modeling analysis give a good account of such an intuitive picture.

\begin{figure}[h]
\begin{center}
\includegraphics[scale=0.3]{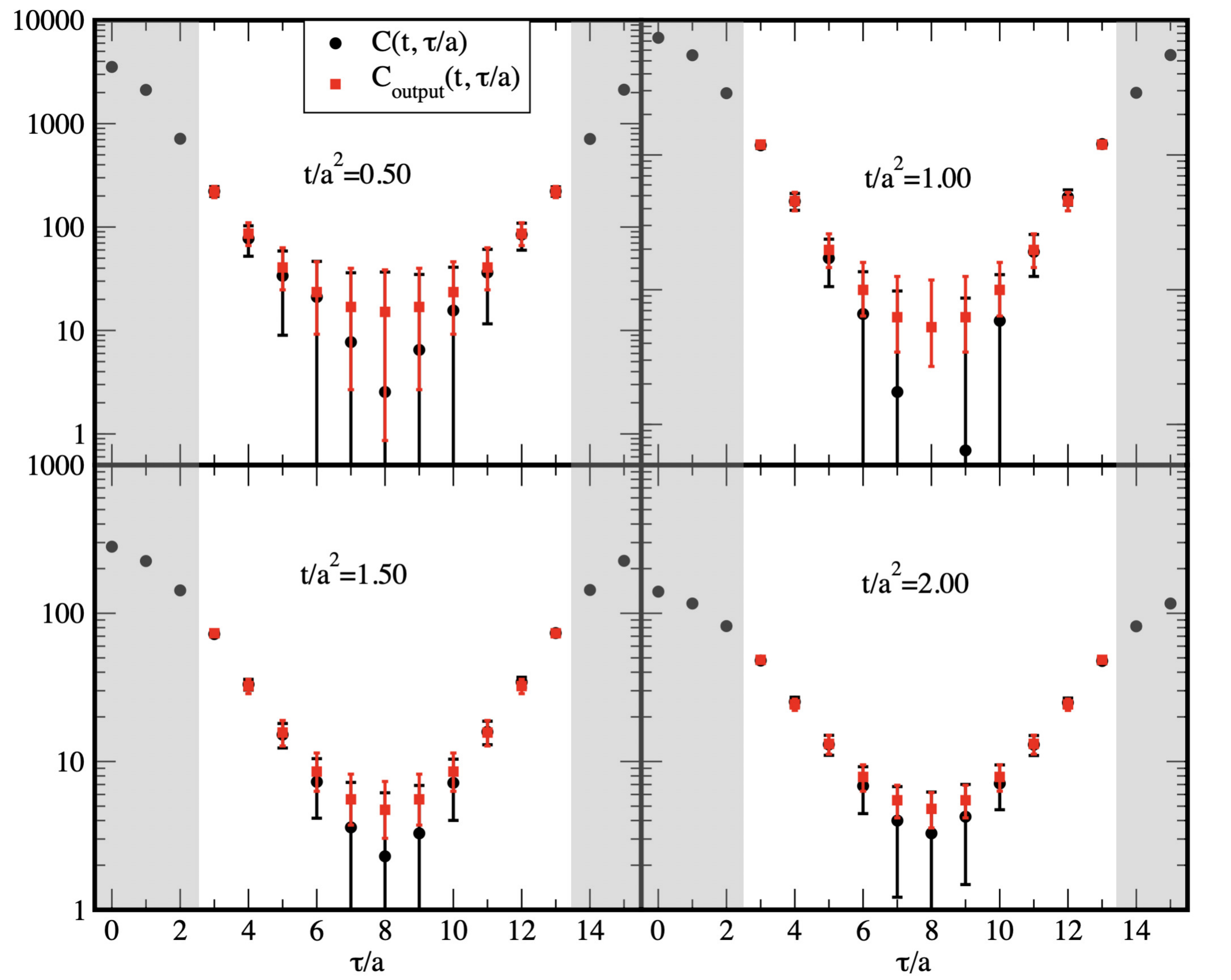}
\caption{The comparison between the input correlation function with the Jackknife error and the output one with the bootstrap error.  The data $C(t,\tau/a)$ in shadowing regime are not utilized in the analyses.}
\label{fig:input-output-Ctau}
\end{center}
\end{figure}

We also show the comparison of the statistical error bars between the input and output $C(t,\tau/a)$ in Fig.~\ref{fig:input-output-Ctau}.
The error bars between them in the short $\tau$-regime are consistent with each other.
The error bars of the output around $\tau/a=N_\tau/2$ are smaller than the ones of the input.
We consider that it comes from the non-negativity condition of the spectral function during the sparse modeling analysis.
It makes $C(t,\tau/a)$ positive correctly, and then the condition reduces the error bars of $C_{output}$.

\subsection*{Acknowledgment}
The work of E.~I. is supported by JSPS KAKENHI with Grant Numbers 
19K03875 and JP18H05407, JSPS Grant-in-Aid for Transformative Research Areas (A) JP21H05190,  JST PRESTO Grant Number JPMJPR2113 and the HPCI-JHPCN System Research Project (Project ID: jh210016).
Numerical simulations were performed on SX-ACE, Octopus and SQUID at the Cybermedia Center, Osaka University.
The calculations were partially performed by the supercomputing system SGI ICE X at the Japan Atomic Energy Agency. This work of Y.~N. was partially supported by JSPS-KAKENHI Grant Numbers 18K11345.


\end{document}